\documentclass[journal,10pt,twocolumn]{IEEEtran}

\usepackage[T1]{fontenc}

\usepackage{graphicx}
\usepackage{amsthm}
\usepackage{soul}
\usepackage{color}

\usepackage{amsmath,cite,amsfonts,amssymb,color, mathtools,setspace,comment,float,caption}
\usepackage{algorithm,algorithmic}
\usepackage{bm}
\usepackage{setspace}
\usepackage{subcaption}
\usepackage{bm}
\usepackage{multirow}
\usepackage{bigstrut}
\DeclarePairedDelimiter\ceil{\lceil}{\rceil}

\newcommand{\be}{\begin{equation}} \newcommand{\ee}{\end{equation}}
\newcommand{\bea}{\begin{eqnarray}} \newcommand{\eea}{\end{eqnarray}}

\usepackage{paracol}
\begin{document}
\title{ Channel Estimation Method and Phase Shift Design for Reconfigurable Intelligent Surface Assisted MIMO Networks}
\author{Jawad Mirza,~\IEEEmembership{Senior Member,~IEEE} and Bakhtiar Ali}
\markboth{}{}%
\maketitle
\begin{abstract}
In this paper, we investigate channel estimation for a large intelligent surface (LIS) aided multiple-input multiple-output (MIMO) communication system. Due to the close proximity of communication devices and LIS terminal, the desired channels can be modeled as line-of-sight channels which are ill-conditioned. To estimate these channel matrices with high quality, we propose a two-stage channel estimation method. In particular, we employ the conventional time-division duplexing based MIMO channel estimation technique in the first stage to estimate the direct MIMO channel between the end terminals. In the second stage, we propose to use a recently developed bilinear adaptive vector approximate message passing (BAdVAMP) algorithm to estimate ill-conditioned LIS channels. The BAdVAMP method has been shown to be accurate and robust for ill-conditioned dictionary learning problems in compressed sensing. We also propose a phase shift design for the LIS using the estimated channels. Specifically, we formulate an optimization problem that maximizes the total channel gain at the user. A closed-form expression to obtain the phase shift of each passive element in the LIS is also derived. Numerical results show that the proposed BAdVAMP based LIS channel estimation performs better than its counterpart scheme bilinear generalized AMP (BiGAMP), especially when channel matrices are ill-conditioned. 
\end{abstract}


\IEEEpeerreviewmaketitle

\section{Introduction}

\IEEEPARstart{T}{he} number of connected mobile devices and the amount of data traffic through these devices are expected to grow many-fold in  future communication networks \cite{cisco2017cisco}. To support the scale of this huge data traffic, more and more base stations and wireless terminals are required to be deployed in existing networks. Nevertheless, practically deploying a large number of base stations having massive antenna arrays will substantially increase the hardware cost and power consumption of the network \cite{7523234}. More recently, large intelligent surface (LIS) aided communication has emerged as a promising reflective-radio technology for next-generation networks capable of fulfilling the demand of high spectral efficiency. It is also considered to be a cost effective and energy efficient solution \cite{basarris}. A typical LIS consists of a planar array having a large number of reflecting metamaterial elements (e.g., low-cost printed dipoles), each of which could act as a phase shift. In particular, the passive or reflective-radio elements of the LIS interact with incident electromagnetic waves and reflect them towards the desired receiver. More importantly, it enables the system to construct a programmable wireless environment suitable for the communication. \cite{ranzoris,LiangLZ0CG19}.

Traditionally, reflecting surfaces have been used in radar and satellite communications with fixed phase shifts. They were not useful in the scenarios where the propagation environment is rapidly changing. However, advancements in RF micro-electromechanical systems (MEMS) and metamaterials have paved the way for reconfigurable reflectors in terrestrial networks, which can be used to enhance received signal power, suppress interference and improve security \cite{8811733}. Many future wireless technologies are being integrated with LIS to investigate the offered performance gain, such as wireless information and power transfer \cite{tang2019joint}, unmanned aerial vehicle (UAV) communication \cite{li2019reconfigurable} and non-orthogonal multiple access (NOMA) \cite{mu2019exploiting}.

Due to its attractive features, the LIS technology has attracted considerable research attention in wireless communication systems employing multiple-input multiple-output (MIMO) technology. The quality of channel state information (CSI) available at the terminals plays an important role in determining the performance of LIS based MIMO systems \cite{8683663,you2019intelligent,zheng2019intelligent,He2019CascadedCE}. In addition to that, an appropriate LIS phase shift design is also important. We briefly discuss some of the existing studies on phase shift design also known as reflect or passive beamforming. For a single-user system, LIS aided MIMO systems are considered in  \cite{abeywickrama2019intelligent,he2019adaptive,wang2019joint} to improve the spectral efficiency, where in \cite{wang2019joint}, a joint phase shift and transmit beamforming vector optimization is carried out at the access point (AP) to maximize the received signal power at the user. In addition to single-user scenarios, the use of LIS has also been investigated for multiuser (MU) scenarios \cite{di2019hybrid,9066923}. According to \cite{di2019hybrid}, the minimum number of transmit antennas required at the AP decreases to half of that required in the traditional hybrid beamforming schemes without LIS, thus, decreasing the hardware cost significantly.

As discussed earlier, the quality of CSI available at the terminals determines the performance of the LIS assisted communication system. Due to this reason, the channel estimation of LIS based channels has attracted an enormous amount of research attention recently \cite{8683663,you2019intelligent,zheng2019intelligent,He2019CascadedCE,wang2019joint,wang2019channel,ning2019channel}. Here, we provide a discussion on the channel estimation work carried out in LIS assisted communication systems. The conventional time-division duplexing (TDD) based MIMO channel estimation method is employed in \cite{8811733} to estimate the AP-user and LIS-user channels. It is assumed that the LIS is equipped with receive RF chains that enable it to estimate the channels. For a single antenna each at AP and user, the channel estimation process in \cite{you2019intelligent} minimizes the mean squared error (MSE) of estimated channels using discrete phases in the training phase. For the same antenna settings as in \cite{you2019intelligent}, the authors in \cite{zheng2019intelligent} employ a cost efficient method with no receive RF chains at the LIS, where concatenated user-LIS-AP channels are estimated at the AP.
Moreover, the idea of grouping LIS elements to reduce the complexity of the channel estimation process is also introduced in \cite{zheng2019intelligent}.

Prior to \cite{zheng2019intelligent}, the estimation of concatenated user-LIS-AP channels was considered in \cite{He2019CascadedCE,8683663}, where ON/OFF switching of LIS elements was implemented. In \cite{8683663}, each passive element controlled by the multi-antenna AP is powered ON one by one to capture the least square estimation of LIS channels. The sparse nature of millimeter wave (mmWave) channels is exploited in \cite{wang2019joint} to convert the channel estimation problem into a sparse signal recovery problem, which is then solved by using compressed sensing algorithms such as orthogonal matching pursuit (OMP) and expectation maximization generalized approximate message passing (EMGAMP). According to \cite{wang2019joint}, the training overhead with the EMGAMP algorithm is lower than the classical OMP algorithm. To reduce the training overhead, a three phase channel estimation process is proposed in \cite{wang2019channel} for MU scenarios, where the correlation among  LIS reflected channels is exploited.

The aforementioned studies on the channel estimation problem either assume a single antenna at both AP and user or multiple antennas at the AP only. In this study, we investigate LIS channel estimation for the system where both AP and user are equipped with multiple antennas. Since, LIS will be densely deployed in indoor and outdoor locations, the near-field communication channels are likely to experience line-of-sight (LOS) propagation conditions \cite{7127536,5360686}. This motivates us to investigate the channel estimation of ill-conditioned\footnote{In an ill-conditioned channel matrix, some of the singular values of the channel are negligible compared to the largest singular value of the channel. The term condition number is generally used to characterize the ill-conditioned channel matrix, which is the ratio of the largest singular value of that channel matrix to the smallest singular value.} LIS channels in MIMO systems. For this purpose, we employ a bilinear adaptive vector approximate message-passing (BAdVAMP) scheme developed in \cite{BAdVAMP} which was originally designed for reconstructing ill-conditioned and mean shifted matrices in dictionary learning problems. This recently developed algorithm has not yet been applied to channel estimation problems in wireless communication systems, therefore, this is the first study to use the BAdVAMP algorithm to estimate LIS based ill-conditioned MIMO channels. Previously, for LIS assisted MIMO systems, the authors in \cite{He2019CascadedCE} have proposed to estimate LIS channels using the approximate message-passing (AMP) based algorithm known as bilinear generalized AMP (BiGAMP). However, the BiGAMP method is designed only for reconstructing well-conditioned matrices, whereas, the BAdVAMP algorithm has shown to be more robust in both ill-conditioned and well-conditioned matrices \cite{BAdVAMP}. Moreover, the BiGAMP method has higher computational complexity and less accuracy as compared to the  BAdVAMP method \cite{BAdVAMP}. 

In this paper, the proposed channel estimation technique is divided into two stages. In the first phase, the direct MIMO channel between user and AP is estimated using the conventional TDD based MIMO channel estimation scheme. Whereas, the BAdVAMP based channel estimation for LIS channels is carried out in the second stage. These estimated channels are used to design the phase shifts for LIS elements. In particular, we derive a closed form expression for phase shifts that maximizes the total channel gain at the user. We assume that a programmable controller is fabricated with LIS which is controlled by the AP to adjust the phases of the LIS elements. The contributions of the paper are summarized below:
\begin{itemize}
    \item We propose a two-stage channel estimation method for the LIS assisted MIMO communication system with multiple transmit and receive antennas at the AP and user, respectively. In the first stage, the direct MIMO channel between the AP and user is estimated, whereas, the LIS channels are estimated in the second stage. The LIS channel matrices are considered to be ill-conditioned, which is a realistic propagation scenario in near-field LOS communications. For the LIS channels, the channel estimation problem is transformed into the well known dictionary learning problem and the recently developed BAdVAMP algorithm is used to learn the ill-conditioned channels.
    \item The LIS channels estimated via the BAdVAMP algorithm have inherent ambiguities out of which the permutation ambiguity has the most destructive effects. Therefore, to get rid of the permutation ambiguity from the recovered channels, we propose an effective method that exploits the sparsity of the phase shift matrix.
    \item Based on the estimated channels, an optimization problem is formulated to obtain the phase shift matrix for the passive elements of LIS which maximizes the total channel gain at the user. Here, we assume a continuous high-resolution phase shift design and derive a closed-form expression to compute the phase shifts of LIS passive elements.
\end{itemize}

The rest of the paper is organized as follows. Section \ref{sys_mod} introduces the LIS assisted MIMO system model. Section \ref{prop} presents the proposed two-stage channel estimation scheme. The ambiguity elimination and the recovery of a unique solution from the estimated channels are discussed in Section \ref{recov}. In Section \ref{p_shift}, we derive the phase shift design for LIS elements. Numerical results are provided in Section \ref{num_sec}. Finally, Section \ref{conc} concludes the paper.

\emph{Notations}: We use $(\cdot)^H$, $(\cdot)^{*}$, $(\cdot)^T$,  $(\cdot)^{-1}$ and $(\cdot)^\perp$ to denote the conjugate transpose, the conjugate, the transpose, the inverse and the pseudoinverse operations, respectively. $\mathbb{E}[\cdot]$ denotes expectation. The Hadamard product is represented by $\odot$. For any given matrix $\mathbf{A}$, the quantity $\mathbf{a}_{i,j}$ denotes the entry of the matrix $\mathbf{A}$ corresponding to the $i^{\text{th}}$ row and $j^{\text{th}}$ column. Similarly, $\mathbf{a}_l$ represents the $l^{\text{th}}$ column of the matrix $\mathbf{A}$.
 
\section{System Model}\label{sys_mod}

\begin{figure}[!t]
    \centering
    \scalebox{0.73}{
   \includegraphics[width=0.6\textwidth]{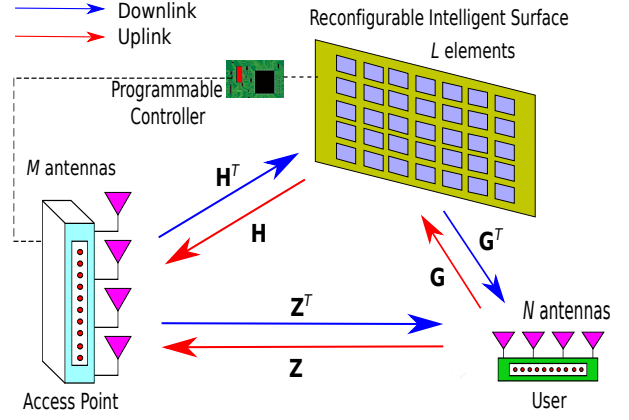}}
    \caption{An illustration of an LIS-enabled MIMO system.}
    \vspace{-1em}
    \label{fig1}
\end{figure}

Consider a single-user MIMO communication system as shown in Fig. \ref{fig1}, where an AP equipped with $M$ antennas serves a user having $N$ antennas. An LIS is installed in a surrounding area, which consists of $L$ low-cost passive reflecting elements. The LIS is connected to the AP via a programmable controller, where the AP is capable of adjusting the phase of passive elements in the LIS terminal. The downlink channels from the AP to the LIS and from the LIS to the user are denoted by $\mathbf{H}^T \in \mathbb{C}^{L \times M}$ and $\mathbf{G}^T \in \mathbb{C}^{N \times L}$, respectively. In this paper, we consider a TDD transmission mode and assume perfect channel reciprocity between the uplink and downlink. Therefore, the uplink channels from the user to the LIS and from the LIS to the AP are given by $\mathbf{G} \in \mathbb{C}^{L \times N}$ and $\mathbf{H} \in \mathbb{C}^{M \times L}$, respectively. The direct downlink and uplink channels between the AP and the user are denoted by $\mathbf{Z}^T \in \mathbb{C}^{M \times N}$ and $\mathbf{Z} \in \mathbb{C}^{N \times M}$, respectively.  

We adopt a narrowband geometric channel model based on extended Saleh-Valenzuela representation to characterize the uplink and downlink channels. Note that the geometric channel model for LIS assisted MIMO/MISO systems has been widely used in previous studies \cite{He2019CascadedCE,wang2019compressed,jamali2019intelligent}. We can express the direct uplink channel between the user and AP as \cite{6717211}
\begin{equation}\label{eq1}
    \mathbf{Z} = \sqrt{\frac{NM}{L_Z}}\sum_{l=1}^{L_{Z}} \alpha_l \mathbf{a}_{\text{r}}\left(\varphi_l^{\text{r}},\vartheta_l^{\text{r}}\right) \mathbf{a}_{\text{t}}^{H}\left(\varphi_l^{\text{t}},\vartheta_l^{\text{t}} \right),
\end{equation}
where $L_Z$ denotes the number of propagation paths and $\alpha_l$ is the complex gain of the $l^{\textrm{th}}$ path for the channel matrix $\mathbf{Z}$. $\mathbf{a}_{\text{r}}\left(\varphi_l^{\text{r}},\vartheta_l^{\text{r}}\right)$ and $\mathbf{a}_{\text{t}}\left(\varphi_l^{\text{t}},\vartheta_l^{\text{t}} \right)$ represent the normalized receive and transmit array response vectors with azimuth (elevation) angles of arrival and departure $\varphi_l^{\text{r}} (\vartheta_l^{\text{r}})$ and $\varphi_l^{\text{t}} (\vartheta_l^{\text{t}})$, respectively. In the case of a uniform planar array (UPA) in the $yz$-plane with $\mathcal{W}$ and $\mathcal{H}$ elements on the $y$ and $z$ axes respectively, an $N=\mathcal{W}\mathcal{H}$ array response vector is
\begin{align}
    \mathbf{a}\left(\varphi,\vartheta\right) = &\sqrt{\frac{1}{N}} \big[ 1, \hdots, e^{jkd\left(m \sin{(\varphi)} \sin{(\vartheta)} + n \cos{(\vartheta)}\right)},  \hdots, \nonumber \\
    & e^{jkd\left(\left(\mathcal{W}-1\right) \sin{(\varphi)} \sin{(\vartheta)} + \left(\mathcal{H}-1\right) \cos{(\vartheta)}\right)} \big]^T,
\end{align}
where $d$ denotes the inter-element spacing. Denoting $\lambda$ as the wavelength, we have $k=2\pi/\lambda$. The complex gain of the $l^{\text{th}}$ path is assumed to be distributed according to independent and identically distributed (i.i.d.) $\mathcal{CN}(0,\sigma^2_{\alpha_{l}})$. The value of $\sigma^2_{\alpha_{l}}$ is set in order to achieve $\mathbb{E}\left[\| \mathbf{Z}\|^2_F \right]=NM$. The same narrowband geometric channel model is assumed for the uplink channels $\mathbf{H}$ and $\mathbf{G}$. For the channel model in \eqref{eq1}, when the number of propagation paths are small (which is the case in high frequency communication links), then the resulting channel is low-rank and also ill-conditioned \cite{5425310}. In addition to this, inadequate antenna spacing at the terminals and small link distance are also key factors that makes the channel ill-conditioned \cite{5454109,jamali2019intelligent}. We assume that all the channels follow a flat quasi-static block fading model\footnote{There are various path loss models developed for LIS-assisted wireless communications \cite{ozdogan2019intelligent}. However, the path loss model does not directly influences the performance of the proposed channel estimation scheme. Therefore, for simplicity, we do not consider effects of the path loss.}. We denote the condition number of the channel matrix $\mathbf{Z}$ by $\kappa(\mathbf{Z})$. Moreover, it is assumed that the channel $\mathbf{G}$ is highly correlated due to a nearly perfect LOS propagation \cite{hu2018beyond,He2019CascadedCE,jung2018performance}, resulting in a low-rank structure of $\mathbf{G}$.

In this study, we consider a \textit{constant amplitude and continuous phase shift} design for reflection coefficients of LIS \cite{LiangLZ0CG19}. Let $s_i \in (0,1)$ and $\phi_i \in (0,2\pi]$ denote the amplitude reflection coefficient (ON/OFF state) and phase shift value at the $i^{\textrm{th}}$ element, respectively, then we can write the phase shift vector for any given time as $\mathbf{s}= [s_1 e^{j\phi_1}, s_2 e^{j\phi_2}, \ldots, s_L e^{j\phi_L}]^T$. Note that the switched off LIS elements may still produce a residual signal, degrading the overall SNR of the signal. However, it is assumed that the ON/OFF operation at the LIS unit is perfect, and therefore, we ignore the imperfections at the LIS unit. For the given $T$ samples or channel uses, the total number of training vectors (channel uses) sent by the user in the first and second stage is denoted by $T_d$ and $T_r$, respectively. Therefore, during the uplink channel estimation process, the received signal at the AP for the first and second stage are
\begin{equation}\label{f_b}
    \mathbf{y}_a\left[t\right] =  \mathbf{Z}\mathbf{x}_a\left[t\right]  + \mathbf{n}_a\left[t\right], \ \ \ t=1,\ldots, T_d 
\end{equation}
and
\begin{equation}\label{f_e}
    \mathbf{y}_b\left[t\right] = \mathbf{H} \left( \mathbf{s}\left[t\right] \odot   \mathbf{G} \mathbf{x}_b\left[t\right]\right) + \mathbf{Z}\mathbf{x}_b\left[t\right]  + \mathbf{n}_b\left[t\right], 
\end{equation}
respectively, where, $\mathbf{x}_a[t] \in \mathbb{C}^{N \times 1}$ and $\mathbf{x}_b[t] \in \mathbb{C}^{N \times 1}$ are the uplink training vectors transmitted by the user in the first and second channel estimation stages, respectively. $\mathbf{n}_a[t]$ and $\mathbf{n}_b[t]$ are the complex additive white Gaussian noise (AWGN) vectors at the AP having zero mean. The covariance matrix of the noise vectors at the first and second stages are given by $\sigma^2_{n_a} \mathbf{I}_M$ and $\sigma^2_{n_b} \mathbf{I}_M$, respectively. We can write the received observation at the AP for the first stage as
\begin{equation}\label{rcv_pilots22}
    \mathbf{Y}_a =  \mathbf{Z} \mathbf{X}_a + \mathbf{N}_a, 
\end{equation}
where, $\mathbf{Y}_a \in \mathbb{C}^{M \times T_d}$, $\mathbf{X}_a=\left[\mathbf{x}_a[1], \mathbf{x}_a[2], \ldots,\mathbf{x}_a[T_d]\right] \in \mathbb{C}^{N \times T_d}$ and $\mathbf{N}_a \in \mathbb{C}^{M \times T_d}$. Similarly, for the second stage of the channel estimation, where the LIS unit is switched ON, we can write the received observation at the AP as
\begin{equation}\label{rcv_pilots}
    \mathbf{Y}_b = \mathbf{H} \left( \mathbf{S} \odot  \mathbf{G} \mathbf{X}_b \right) + \mathbf{Z} \mathbf{X}_b+ \mathbf{N}_b, 
\end{equation}
where, $\mathbf{Y}_b \in \mathbb{C}^{M \times T_r}$, $\mathbf{X}_b=\left[\mathbf{x}_b[T_d+1], \mathbf{x}_b[T_d+2], \ldots,\mathbf{x}_b[T_d+T_r]\right] \in \mathbb{C}^{N \times T_r}$, $\mathbf{S}=\left[\mathbf{s}[T_d+1], \mathbf{s}[T_d+2], \ldots, \mathbf{s}[T_d+T_r]\right]  \in \mathbb{C}^{L \times T_r}$ and $\mathbf{N}_b \in \mathbb{C}^{M \times T_r}$. After the channel estimation process, the downlink received signal at the user is given by
\begin{equation}
    \mathbf{r} = \left(\mathbf{G}^T \mathbf{\Phi} \mathbf{H}^T + \mathbf{Z}^T \right) \mathbf{W}\mathbf{u} + \mathbf{n},
\end{equation}
where, $\mathbf{\Phi}= \text{diag}\left(s_1 e^{j\phi_1}, s_2 e^{j\phi_2}, \ldots, s_L e^{j\phi_L}\right)$ is the phase shift matrix. The precoding matrix at the AP is given by $\mathbf{W} \in \mathbb{C}^{M \times N_s}$, such that $\|\mathbf{W}\|^2_F = N_s$, where the total number of data streams is denoted by $N_s$. The data vector is $\mathbf{u} \in \mathbb{C}^{N_s \times 1}$ and $\mathbf{n}\sim\mathcal{CN}(0,\sigma^2_{n} \mathbf{I}_N)$ is an AWGN vector for the user. Denoting $\check{\mathbf{G}}^T$, $\check{\mathbf{H}}^T$ and $\check{\mathbf{Z}}^T$ as the estimates of $\mathbf{G}^T$, $\mathbf{H}^T$ and $\mathbf{Z}^T$, respectively, the transmit precoding matrix, $\mathbf{W}$, is equal to the right singular vectors of the composite estimated channel matrix $\left(\check{\mathbf{G}}^T \mathbf{\Phi} \check{\mathbf{H}}^T + \check{\mathbf{Z}}^T \right)$ (obtained via a singular value decomposition), corresponding to the $N_s$ largest singular values. The user estimates the downlink channel during the downlink training phase and uses the left singular vectors of the estimated composite channel corresponding to the $N_s$ largest singular values, and uses them as a receive combining matrix, $\mathbf{V} \in \mathbb{C}^{N \times N_s}$, such that $\|\mathbf{V}\|^2_F = N_s$. This is known as eigenmode transmission as it maximizes the signal-to-noise ratio (SNR) and achieves full diversity order \cite{paulraj2003introduction}. The received signal at the user can be written as
\begin{equation}
    \tilde{\mathbf{r}} = \mathbf{V}^H \left(\mathbf{G}^T \mathbf{\Phi} \mathbf{H}^T + \mathbf{Z}^T \right) \mathbf{W} \mathbf{u} + \mathbf{V}^H\mathbf{n}.
\end{equation}
The achievable rate of the system is given by
\begin{align}\label{cap}
    C &= \log_2 \text{det} \left( \mathbf{I} + \rho \left( \mathbf{W}^H\bm{\Delta}^H \mathbf{V}\mathbf{V}^H \bm{\Delta} \mathbf{W}\right) \right), 
\end{align}
where $\bm{\Delta}=\mathbf{G}^T \mathbf{\Phi} \mathbf{H}^T + \mathbf{Z}^T$ and SNR$=\rho \overset{\Delta}{=} 1/\sigma^2_{n}$. Estimated channels are used to design the phase shift matrix, $\bm{\Phi}$. 

\section{Proposed Two-Stage Channel Estimation} \label{prop}
In this section, we present our proposed two stage channel estimation method to estimate the three channels: $\mathbf{Z}$, $\mathbf{H}$ and  $\mathbf{G}$. In the first stage, the direct uplink MIMO channel from the user to the AP, given by $\mathbf{Z}$, is estimated using the conventional training-based MIMO channel estimation technique \cite{1597555}. Whereas, in the second stage, the LIS channels $\mathbf{H}$ and $\mathbf{G}$ are estimated at the AP using the BAdVAMP scheme \cite{BAdVAMP}. Here, we only provide the details of the uplink channel estimation at the AP via uplink training vectors from the user. A similar procedure can be adapted at the user to estimate the downlink channels $\mathbf{H}^T$, $\mathbf{G}^T$ and $\mathbf{Z}^T$. However, we skip the details of the downlink channel estimation process. 

\subsection{Conventional TDD based MIMO channel estimation}
In the first stage of the channel estimation procedure at the AP, the LIS terminal is powered off, i.e., $\{s_i\} = 0$, $\forall i$, and the AP estimates the direct uplink channel $\mathbf{Z}$. In this stage, $T_d$ training vectors are transmitted by the user, such that $T_d \geq N$. We adopt relaxed MMSE (RMMSE) channel estimation at the AP, presented in \cite{1597555}. Using the RMMSE criterion, the estimated direct channel can be expressed as 
\begin{equation}
    \check{\mathbf{Z}} =  \mathbf{Y}_a  \left(\mathbf{X}_a^H \mathbf{X}_a + \sigma_{n_a}^2 \mathbf{I}_{T_d}\right)^{-1} \mathbf{X}_a^H  ,
\end{equation}
where the training matrix satisfies $\| \mathbf{X}_a\|_F^2 = p_u$. Similar to \cite{1597555}, we rely on discrete Fourier transform (DFT) matrix design for $\mathbf{X}_a$, given by
\begin{equation}
    \mathbf{X}_a = \sqrt{\frac{p_u}{T_d N}} \begin{bmatrix} 1 & 1 & \hdots & 1 \\ 
    1 & \text{e}^{\frac{j2\pi}{T_d}} & \hdots & \text{e}^{\frac{j2\pi(T_d-1)}{T_d}} \\
    1 & \vdots & \ddots & \vdots \\
    1 & \text{e}^{\frac{j2\pi(N-1)}{T_d}} & \hdots & \text{e}^{\frac{j2\pi(N-1)(T_d-1)}{T_d}}
    \end{bmatrix}\end{equation}
Note that in the RMMSE channel estimation scheme a prior knowledge of the channel parameters are not needed, however, only the receiver noise
power is required. 

\subsection{BAdVAMP based channel estimation} \label{bvamp}
After the estimation of the direct MIMO channel between the AP and user, the second stage of the channel estimation process begins, where the user sends $T_r$ training vectors to the AP. In this stage, the LIS terminal is switched ON and the received signal at the AP is given by \eqref{f_e}. The AP removes the direct channel part from \eqref{rcv_pilots} by using $\check{\mathbf{Z}}$ and $\mathbf{X}_b$, such that the resulting observation for all $T_r$ training vectors at the AP becomes
\begin{equation}
        \mathbf{Y} = \mathbf{H} \left( \mathbf{S} \odot  \mathbf{G} \mathbf{X}_b \right) + \mathbf{N}, 
        \label{eq_8}
\end{equation}
where $\mathbf{N}$ is the noise matrix which is given by $\mathbf{N}=\mathbf{N}_b + \mathbf{Z}\mathbf{X}_b - \check{\mathbf{Z}}\mathbf{X}_b$. Therefore, the entries of $\mathbf{N}$ follow the $\mathcal{CN}(0,\sigma_{n_b}^2+\sigma_e^2)$ distribution, where $\sigma_{n_b}^2$ and $\sigma_e^2$ denote the noise and error variances, respectively. The phase shift design at the LIS unit also plays an important role during the channel estimation process. For the second channel estimation stage, we assume that ON/OFF state values $\{s_i\}$ of LIS passive elements are independent and for the given training duration $t$, $\mathbf{s}[t]$ is a $K$-sparse vector, such that total $K$ non-zero value are in the vector $\mathbf{s}[t]$. Note that the sparse nature of the matrix $\mathbf{S}$ in \eqref{eq_8} is necessary for channel estimation via the BAdVAMP algorithm. It is also assumed that $\mathbf{X}_b$ is a full rank matrix and $\|\mathbf{X}_b)\|^2_F = p_u$.

We propose to use the BAdVAMP algorithm to estimate channels $\mathbf{H}$ and $\mathbf{G}$ from \eqref{eq_8}. The BAdVAMP algorithm is known to be a computationally efficient method for estimating the matrices $\mathbf{H}$ and $\mathbf{D}$ from the noisy signal $\mathbf{Y} = \mathbf{H} \mathbf{D} + \mathbf{N}$ where, $\mathbf{D} = \mathbf{S} \odot  \mathbf{G} \mathbf{X}_b$. This is also known as the dictionary learning problem in the domain of compressive sensing \cite{parker2014bilinear,BAdVAMP}. The BAdVAMP algorithm \cite{BAdVAMP} was shown to be superior to its contenders both in complexity and performance, especially when the matrices $\mathbf{H}$ and $\mathbf{D}$ are ill-conditioned.

We aim to estimate the matrix $\mathbf{D}\in\mathbb{C}^{L\times T_r}$, from the noisy received matrix $\mathbf{Y}\in\mathbb{C}^{M\times T_r}$ in \eqref{eq_8}. As we are dealing with the scenario where $L > M$ and $\mathbf{D}$ is a sparse matrix, our channel estimation problem becomes equivalent to the estimation of sparse vectors from underdetermined linear measurements. In addition, as the matrix $\mathbf{H}$ is also not known, the problem becomes a bilinear recovery problem. We can express the channel matrix $\mathbf{H}$ as an unstructured matrix \cite{BAdVAMP}
\begin{equation}
    \mathbf{H}(\bm{\psi}_{H}) = \sum_{m=1}^{M}\sum_{l=1}^{L}\psi_{H,m,l}\mathbf{e}_m\mathbf{e}_l^T,
\end{equation}
where, $\mathbf{e}_m$ denotes the $m^{\text{th}}$ basis vector and $\bm{\psi}_H\in \mathbb{C}^{M\times L}$. The vector $\mathbf{e}_m$ will have a value one at the $m^{\text{th}}$ location and zero elsewhere. We assume that the density of $\mathbf{D}$, given by $p_{\mathbf{D}}(.;\bm{\theta}_{d})$, is parameterized by the vector $\bm{\theta}_{d}$. More specifically, the $n^{\text{th}}$ value of the vector $\bm{\theta}_{d}$ represents the mean and variance values of the $n^{\text{th}}$ column of $\mathbf{D}$. In order to estimate $\mathbf{H}$ and $\mathbf{D}$, the BAdVAMP aims to learn the parameters $\mathbf{\Theta}\triangleq \{\bm{\psi}_{H}, \bm{\theta}_{d} \}$ from $\mathbf{Y}$. In this study, it is assumed that the statistical information $\bm{\theta}_{d}$ is known at the AP and we focus on recovering $\bm{\psi}_{H}$ and $\mathbf{D}$ from the noisy measurement matrix $\mathbf{Y}$ in \eqref{eq_8}, which can be rewritten as 
\begin{equation}\label{Stmodel}
    \mathbf{Y} = \mathbf{H}(\bm{\psi}_{H}) \mathbf{D} + \mathbf{N}.
\end{equation}

\subsubsection{Decoupling the Estimation Problem}

To compute the parameters in $\bm{\Theta}$ that maximize the received signal distribution, the BAdVAMP algorithm uses maximum likelihood (ML) estimation given by 
\begin{equation}\label{Eq_ML}
    \Hat{\bm{\Theta}}_{\text{ML}} = \text{arg}\max_{\bm{\Theta}} \ \ p_{\mathbf{Y}}(\mathbf{Y};\bm{\Theta}).
\end{equation}
Based on these estimated parameters, the MMSE estimate of $\mathbf{D}$ can be expressed as \cite{BAdVAMP}
\begin{equation} \label{Eq_MMSE}  
    \Hat{\mathbf{D}}_{\text{MMSE}} = \mathbb{E}\left[\mathbf{D}|\mathbf{Y};\hat{\bm{\Theta}}_{\text{ML}}\right].
\end{equation}
The BAdVAMP algorithm approximates the quantities in \eqref{Eq_ML} and \eqref{Eq_MMSE}, i.e., $\Hat{\bm{\Theta}}_{\text{ML}}$ and $\Hat{\mathbf{D}}_{\text{MMSE}}$, respectively. To find the expected value in \eqref{Eq_MMSE}, the mean of the posterior density $ p_{\mathbf{D}|\mathbf{Y}}(\mathbf{D}|\mathbf{Y};\hat{\bm{\Theta}}_{\text{ML}})$ is required. Using Bayes' rule we can write the posterior density as
\begin{equation} \label{pd}
    p_{\mathbf{D}|\mathbf{Y}}(\mathbf{D}|\mathbf{Y};\hat{\bm{\Theta}}_{\text{ML}})=\frac{p_{\mathbf{Y}|\mathbf{D}}(\mathbf{Y}|\mathbf{D};
    \hat{\bm{\Theta}}_{\text{ML}})p_{\mathbf{D}}(\mathbf{D};\hat{\bm{\Theta}}_{\text{ML}})}{p_{\mathbf{Y}}(\mathbf{Y};\hat{\bm{\Theta}}_{\text{ML}})}
\end{equation}
where $p_{\mathbf{Y}|\mathbf{D}} (\mathbf{Y}|\mathbf{D};\hat{\bm{\Theta}}_{\text{ML}})$ is the likelihood function of \eqref{Eq_MMSE}, $p_{\mathbf{D}}(\mathbf{D};\hat{\bm{\Theta}}_{\text{ML}})$ denotes the prior density and the likelihood function of \eqref{Eq_ML} is represented by $p_{\mathbf{Y}}(\mathbf{Y};\hat{\bm{\Theta}}_{\text{ML}})$. Following the decoupling of the posterior density \eqref{pd} across the columns of $\mathbf{D}$, as in \cite{BAdVAMP}, we can write $p_{\mathbf{D}}(\mathbf{D};\hat{\bm{\Theta}}_{\text{ML}})=\prod_{t=1}^{T_r}p_{\mathbf{d}}(\mathbf{d}_{t};{\bm{\theta}}_{d})
$ and $p_{\mathbf{Y}|\mathbf{D}}(\mathbf{Y}|\mathbf{D};\hat{\bm{\Theta}}_{\text{ML}})=\prod_{t=1}^{T_r}p_{\mathbf{y}|\mathbf{d}}(\mathbf{y}_{t}|\mathbf{d}_{t};\hat{\bm{\Theta}}_{\text{ML}})$. Now, we can express \eqref{pd} as
\begin{equation}
p_{\mathbf{D}|\mathbf{Y}}(\mathbf{D}|\mathbf{Y};\hat{\bm{\Theta}}_{\text{ML}})\hspace{-.2em} \propto\hspace{-.2em} \prod_{t=1}^{T_r} p_{\mathbf{d}}(\mathbf{d}_{t};\hat{\bm{\Theta}}_{\text{ML}}) p_{\mathbf{y}|\mathbf{d}}(\mathbf{y}_{t}|\mathbf{d}_{t};\hat{\bm{\Theta}}_{\text{ML}})
\end{equation}
Using this decoupled posterior density, the BAdVAMP algorithm uses a vector AMP (VAMP) technique \cite{rangan2019vector} to find the matrix $\mathbf{D}$ from \eqref{Stmodel}, where each column of $\mathbf{D}$ is tracked independently. However, the VAMP algorithm requires complete information of $\bm{\Theta}$ to estimate $\mathbf{D}$. From \eqref{Eq_ML} and \eqref{pd}, the ML estimate of $\bm{\Theta}$ can be computed as
\begin{equation}\label{n_p}
    \hat{\bm{\Theta}}_{\text{ML}} = \text{arg}\min_{\bm{\Theta}} \ - \ln \int p_{\mathbf{D}}(\mathbf{D};{\bm{\Theta}}) p_{\mathbf{Y}|\mathbf{D}}(\mathbf{Y}|\mathbf{D};{\bm{\Theta}}) \ \text{d}\mathbf{D}.
\end{equation}
The BAdVAMP algorithm uses the expectation maximization (EM) approach \cite{dempster1977maximum} to solve the problem \eqref{n_p} in an iterative manner. Therefore, the BAdVAMP algorithm interleaves the EM and VAMP algorithms to estimate $\mathbf{H}$ and $\mathbf{D}$ from $\mathbf{Y}$.

\subsubsection{The BAdVAMP Algorithm}
Here, we discuss the BAdVAMP algorithm \cite{BAdVAMP} to estimate $\mathbf{H}$ and $\mathbf{D}$. As discussed earlier, the BAdVAMP algorithm employs the EM approach to estimate $\bm{\psi}_{H}$ and the VAMP \cite{7869633} algorithm to estimate $\mathbf{D}$. In particular, the BAdVAMP method\footnote{Here, we present the brief overview of the BAdVAMP algorithm. We refer the reader to \cite{BAdVAMP} for a detailed description of the algorithm.} performs a joint estimation of parameters using the optimization problem based on the Gibbs free energy \cite[Eq. (38)]{BAdVAMP}. We can divide the BAdVAMP quantities into two groups: VAMP and EM quantities. The VAMP quantities are $\{\mathbf{r}_1,\gamma_1,\mathbf{r}_2,\gamma_2,\hat{\mathbf{d}},\eta\}$ and $\bm{\Theta}$ is the EM quantity. These quantities are computed and are updated iteratively by using Algorithm~\ref{alg}. In Algorithm 1, $\mathbf{g}_1(\mathbf{r}_{1,t},\gamma_{1,t};\bm{\theta}_{d})$ represents a denoising function given by 
\begin{equation}
    \mathbf{g}_1(\mathbf{r}_{1,t},\gamma_{1,t};\bm{\theta}_{d}) \triangleq \frac{\int \mathbf{d}_t p_{\mathbf{d}} (\mathbf{d}_t;\bm{\theta}_{{d}}) \bm{\xi}(\mathbf{d}_t)\text{d}\mathbf{d}_t} {\int p_{\mathbf{d}} (\mathbf{d}_t;\bm{\theta}_{{d}})\bm{\xi}(\mathbf{d}_t)\text{d}\mathbf{d}_t},
\end{equation}
where $\bm{\xi}(\mathbf{d}_t) \sim \mathcal{CN}(\mathbf{d}_t;\mathbf{r}_{1,t},\mathbf{I}/\gamma_{1,t})$ and $t$ represents the column of the matrix $\mathbf{D}$. In Algorithm \ref{alg}, the quantity $\langle \mathbf{g}'_1(\mathbf{r}_{1,t},\gamma_{1,t},\bm{\theta}_{d})\rangle$ represents the divergence of $\mathbf{g}_1(\mathbf{r}_{1,t},\gamma_{1,t};\bm{\theta}_{d})$ at $\mathbf{r}_{1,t}$. In general, $\mathbf{g}_1(\mathbf{r}_{1,t},\gamma_{1,t};\bm{\theta}_{{d}})$ can be interpreted as denoising of the AWGN-corrupted pseudo-measurement $\mathbf{r}_{1,t} = \mathbf{d}_{1,t}+\mathbf{f}_{1,t}$, where $\mathbf{f}_{1,t}\sim\mathcal{CN}(0,\mathbf{I}/\gamma_{1,t})$, which corresponds to the first loop i.e., $\tau_1$ in Algorithm~\ref{alg}, from line 3 till line 7. On the other hand, the second loop $\tau_2$ performs MMSE estimation of $\mathbf{d}_{2,t}$ from the AWGN-corrupted  measurements under the pseudo-prior $\mathbf{d}_{2,t}\sim\mathcal{CN}(\mathbf{r}_{2,t},\mathbf{I}/\gamma_{2,t})$. Lastly, the estimate of $\mathbf{H}(\bm{\psi}_{H}^t)$ is given in lines 10-16. The final updated estimate of $\mathbf{H}$ using Algorithm~\ref{alg} is given by 
\begin{equation}
    \hat{\mathbf{H}} = \mathbf{H}(\bm{\psi}_{{H}}^{i}) = \mathbf{Y}\mathbf{D}_2^{i^T} (\mathbf{C}^i + \mathbf{D}_2^i \mathbf{D}_2^{i^T})^{-1},
\end{equation}
where $\mathbf{D}_2^{i}=\left[{\mathbf{d}_{2,1}^i},\mathbf{d}_{2,2}^i,\dotsc, {\mathbf{d}_{2,T_r}^i}\right]$ and $i=T_{\text{max}}$. On the other hand, the estimate of $\hat{\mathbf{D}}$ is given by
\begin{equation}
    \hat{\mathbf{D}} = \mathbf{D}_2^{i} = \left[{\mathbf{d}_{2,1}^i},\mathbf{d}_{2,2}^i,\dotsc, {\mathbf{d}_{2,T_r}^i}\right],
\end{equation}
where $i=T_{\text{max}}$.
\begin{algorithm}[!t]       \label{aa1}               
\caption{Bilinear Adaptive VAMP (BAdVAMP)}          
\label{alg}                           
\begin{algorithmic}[1]                    
  \STATE {{\textbf{Initialization}}:} \\ \hspace{4mm} $\forall$ : $\mathbf{r}_{1,t}^0, \gamma_{1,t}^0, \bm{\theta}_{d}, \bm{\psi}_{H}^0$
     \STATE \textbf{for} $ i=0,\dotsc,T_{\max}$ \textbf{do}
        \STATE \hspace{4mm} \textbf{for} $\tau_1=0,\dotsc,\tau_{1,\max}$ \textbf{do} 
            \STATE \hspace{8mm} $\forall t$ : $\mathbf{d}_{1,t}^i \gets \mathbf{g}_1(\mathbf{r}_{1,t}^i,\gamma_{1,t}^i,\bm{\theta}_{d})$
            \STATE \hspace{8mm} $\forall t$ : $1/\eta_{1,t}^i \gets\langle \mathbf{g}'_1(\mathbf{r}_{1,t}^i,\gamma_{1,t}^i,\bm{\theta}_{d})\rangle/\gamma_{1,t}^i$ 
            \STATE \hspace{8mm} $\forall t$ :  $1/\gamma_{1,t}^i \gets \frac{1}{N}\|\mathbf{d}_{1,t}^i-\mathbf{r}_{1,t}^i\|^2 + 1/\eta_{1,t}^i$
        \STATE \hspace{4mm} \textbf{end for}
        \STATE \hspace{4mm} $\forall t$ : $\gamma_{2,t}^i = \eta_{1,t}^i - \gamma_{1,t}^i$
        \STATE \hspace{4mm} $\forall t$ : $\mathbf{r}_{2,t}^i = (\eta_{1,t}^i\mathbf{d}_{1,t}^i - \gamma_{1,t}^i\mathbf{r}_{1,t}^i)/\gamma_{2,t}^i$
        \STATE \hspace{4mm} \textbf{for} $\tau_2=0,\dotsc,\tau_{2,\max}$ \textbf{do}
                \STATE \hspace{8mm} $\forall t$ : $\mathbf{C}_{t}^i\gets (\gamma_{2,t}^i\mathbf{I}_{N} + {\mathbf{H}(\bm{\psi}}_{H}^i)^T\mathbf{H}(\bm{\psi}_{H}^i))^{-1}$
            \STATE \hspace{8mm} $\forall t$ : $\mathbf{d}_{2,t}^i \gets \mathbf{C}_l^i(\gamma_{2,t}^i\mathbf{r}_{2,t}^i + \mathbf{H}(\bm{\psi}_{H}^i)^T\mathbf{y}_{t})$
            \STATE \hspace{8mm} $\forall t$ : $1/\eta_{2,t}^i \gets \text{tr}\{\mathbf{C}_{t}^i\}/N$
            \STATE \hspace{8mm} $\mathbf{C}^i \gets \sum_{t=1}^{T_r}\mathbf{C}_{t}^i$
            \STATE \hspace{8mm} $\mathbf{H}(\bm{\psi}_{H}^i) \gets \mathbf{Y}{\mathbf{D}_2^{i^T}} ({\mathbf{C}^i} + {\mathbf{D}_2^i} \mathbf{D}_2^{i^T})^{-1}$
        \STATE \hspace{4mm} \textbf{end for}
        \STATE \hspace{4mm} $\mathbf{H}(\bm{\psi}_{H}^{t+1})=\mathbf{H}(\bm{\psi}_{H}^t)$
        \STATE \hspace{4mm} $\forall t$ : $\gamma_{1,t}^{i+1} = (\eta_{2,t}^i-\gamma_{2,t}^i)$
        \STATE \hspace{4mm} $\forall t$ : $\mathbf{r}_{1,t}^{i+1} = (\eta_{2,t}^i\mathbf{d}_{2,t}^i - \gamma_{2,t}^i\mathbf{r}_{2,t}^i)/\gamma_{1,t}^{i+1}$
    \STATE \textbf{end for}
            
\end{algorithmic}
\end{algorithm}
The BAdVAMP algorithm presented in Algorithm~\ref{alg} recovers $\hat{\mathbf{H}}$ and $\hat{\mathbf{D}}$ up-to certain phase, scalar and permutation ambiguities. For a unique solution, it is important to remove the ambiguities from $\hat{\mathbf{H}}$ and $\hat{\mathbf{D}}$.

\section{Removing Permutation Ambiguity and Recovery of $\mathbf{G}$}\label{recov}

The estimated channel matrices obtained via the BAdVAMP algorithm contain ambiguities, which can be classified into three main categories: permutation, scalar and phase ambiguities. It is not possible to avoid inherent scalar and phase ambiguities for the studied system model. However, by leveraging the sparsity in $\mathbf{D}$, the permutation ambiguity can be removed from the estimated matrices $\hat{\mathbf{H}}$ and $\hat{\mathbf{D}}$. Here, we also discuss a low-rank matrix completion approach to recover $\mathbf{G}$ after removing the permutation ambiguity from $\hat{\mathbf{D}}$. We begin by explaining the ambiguity issue that if $\hat{{\mathbf{H}}}$  and $\hat{{\mathbf{D}}}$ represent the estimated channel matrices obtained via the BAdVAMP algorithm, then $\hat{{\mathbf{H}}} \mathbf{\Sigma} \mathbf{\Gamma} $ and $\mathbf{\Gamma}^T \mathbf{\Sigma}^{-1} \hat{{\mathbf{D}}} $ are also  valid solutions for any arbitrary permutation matrix $\mathbf{\Gamma}$ and diagonal matrix $\mathbf{\Sigma}$, where $\mathbf{\Sigma}$ comprises of scalar and phase ambiguities. 
\subsection{Removing Permutation Ambiguity}
The design of phase shifters $\mathbf{S}$  during the training period is crucial for removing the permutation ambiguity that is induced in the estimated channels. As each column of $\mathbf{S}$ is a $K-$sparse vector, we define a state matrix for the matrix $\hat{{\mathbf{D}}}$ as
\begin{equation}\label{s_m}
    \bar{\mathbf{S}}_{l,t} \overset{\Delta}{=} \begin{cases} 
    1,\ \ \ \hat{{\mathbf{D}}}_{l,t} \neq 0 \\
    0,\ \ \ \hat{{\mathbf{D}}}_{l,t} = 0
    \end{cases}
\end{equation}
From \eqref{s_m}, the relationship between $\bar{\mathbf{S}}$ and ${\mathbf{S}}$, can be expressed as $\bar{\mathbf{S}} = {\mathbf{\Gamma}} {\mathbf{S}}$. This relationship shows that the $n^{\text{th}}$ row of the state matrix, denoted by $\bar{\mathbf{s}}^{T}_n$, represents the $n{'}$ row of $\mathbf{S}$ given by ${\mathbf{s}}^T_{n{'}}$. This suggests that the $(n,n^{'})^{\text{th}}$ element of the permutation matrix $\mathbf{\Gamma}$ must be equal to one, i.e., ${\mathbf{\Gamma}}_{n,n^{'}} = 1$. Using this fact, we can compute the $n^{\text{th}}$ row of ${\mathbf{\Gamma}}$ with a non-zero value at the location $\hat{n}$, where
\begin{equation}\label{opt_max}
    \hat{n} = \underset{n^{'}}{\mathrm{argmax}} \left(  {{\mathbf{s}}_{n{'}} \bar{\mathbf{s}}}_n^T \right).
\end{equation}
This approach yields the permutation ambiguity matrix ${{\mathbf{\Gamma}}}$ which helps in removing the permutation ambiguity from the BAdVAMP estimated channels, such that the recovered channel matrices are given by $\check{\mathbf{D}}={{\mathbf{\Gamma}}}^T \hat{{\mathbf{D}}}$ and 
$\check{\mathbf{H}}= \hat{{\mathbf{H}}}{{\mathbf{\Gamma}}}$. 

\subsection{Estimation of $\mathbf{G}$ via Matrix Completion}

To extract the estimate of the channel matrix $\mathbf{G}$ from the permutation ambiguity removed matrix $\check{\mathbf{D}}$, similar to \cite{He2019CascadedCE}, we use a low-rank matrix completion approach. The main idea in the low-rank matrix completion approach is to recover a low-rank matrix from the set of linear measurements. In this study, we use a well-known normalized iterative hard thresholding (NIHT) \cite{tanner2013normalized} algorithm to recover $\mathbf{G}$ from the matrix $\check{\mathbf{D}}$. The NIHT algorithm works on the principle of projection based gradient descent. 

In a matrix completion approach, if a recoverable matrix of size $L \times N$ has a rank of $r$, then $r(L+N-r)$ measurements may be sufficient for recovery \cite{tanner2013normalized}. In our problem, we want to recover $\mathbf{G}$ from the estimated matrix $\check{\mathbf{D}}$. Therefore, we can write the matrix completion problem as
\begin{align}
    &\min_{\mathbf{F}} \frac{1}{2} \| \check{\mathbf{D}} - \left( \mathbf{S} \odot \mathbf{F}\right)\|^2_F \\
    & \text{s. t.} \ \ \ \ \ \text{rank}(\mathbf{F}) = r_g.
\end{align}
where $r_g$ denotes the rank of the matrix $\mathbf{G}$. The NIHT based method to find the estimate of $\mathbf{G}$ is presented in Algorithm 2. 
\begin{algorithm}[!t]       \label{aa2}               
\caption{NIHT based method to recover $\mathbf{G}$}          
\label{alg2}                           
\setstretch{1.15}
\begin{algorithmic}[1]                    
\STATE{\textbf{Input}}:  $r_g$, $\check{\mathbf{D}}$, $\mathbf{S}$ and $\mathbf{X}_b$
     \STATE \textbf{Set}: $\mathbf{F}^{0} = \mathbf{0}, j=0$ and $\mathbf{U}^0$ (top $r_g$ left singular vectors of $\mathbf{F}^{0}$)
        \STATE \hspace{4mm} \textbf{Repeat} 
            \STATE \hspace{8mm} \textbf{Set} $\mathbf{P}_{U}^j = \mathbf{U}^j \mathbf{U}^{j^{H}}$ 
            \STATE \hspace{8mm} \textbf{Compute} $\alpha^j = \frac{\| \mathbf{P}_U^j \mathbf{S} \odot \left(\check{\mathbf{D}} - \mathbf{F}^j \right)\|^2_F}
            {\| \mathbf{S} \odot \left( \mathbf{P}_U^j \mathbf{S} \odot \left(\check{\mathbf{D}} - \mathbf{F}^j \right) \right)\|^2_2} $  
            \STATE \hspace{8mm} \textbf{Set} $\mathbf{Q} = \mathbf{F}^j + \alpha^j \mathbf{P}_U^j \mathbf{S} \odot \left( \check{\mathbf{D}} -  \mathbf{F}^{j}\right)$
            \STATE \hspace{8mm} \textbf{Set} $\mathbf{F}^{j+1} = \mathcal{H}_{r_g} \left( \mathbf{Q}\right)$
        \STATE \hspace{8mm} \textbf{Set} $\mathbf{U}^{j+1}$ to top $r_g$ left singular vectors of $\mathbf{F}^{j+1}$
        \STATE \hspace{8mm} $j=j+1$
        \STATE \hspace{4mm} \textbf{Until} $j >$ max$\_$iter
        \STATE \textbf{Output} $\check{\mathbf{G}} = \mathbf{F}^j (\mathbf{X}_b\mathbf{X}_b^H)^{-1}\mathbf{X}_b$ 
\end{algorithmic}
\end{algorithm}
In Algorithm \ref{alg2}, the operation $\mathcal{H}_{r_g}(\cdot)$ restricts the rank of the input to the $r_g$ by performing SVD on the input and keeping only $r_g$ singular values in the diagonal singular value matrix, while replacing the remaining singular values with zero.

Now we have estimated the LIS channels and the direct channel which are $\{\check{\mathbf{H}},\check{\mathbf{G}}\}$ and $\check{\mathbf{Z}}$, respectively. Note that scalar and phase ambiguities are still present in the LIS estimated channels. However, the precoding/combining strategy and LIS phase shift design considered in this study depend on the composite channel given by $\check{\mathbf{H}} \mathbf{\Phi} \check{\mathbf{G}} + \check{\mathbf{Z}}$. Therefore, there is no need to eliminate scalar and phase ambiguities. To give more insight into this, let us assume that 
$\check{\mathbf{H}}'$ and $\check{\mathbf{G}}'$ denote the estimated channels with perfect elimination of the phase and scalar ambiguities, then we can write $\check{\mathbf{H}} \mathbf{\Phi} \check{\mathbf{G}} + \check{\mathbf{Z}}=\check{\mathbf{H}}' \mathbf{\Phi} \check{\mathbf{G}}' + \check{\mathbf{Z}}$ \cite{He2019CascadedCE}. Therefore, it is sufficient to have only permutation ambiguity removed from the estimated channels to design the LIS phase shifts and MIMO precoding/combining vectors. 

\section{Optimization of LIS Phase Shifts} \label{p_shift}
Phase shifts of all passive elements in the LIS play an important role in determining the performance of the LIS assisted MIMO systems. In particular, phase shifts of passive elements in the LIS can be adjusted according to the propagation environment such that the reflected signals add up coherently at the user with the signals arriving from other paths. This adjustment helps improve the total channel gain or SNR of the received signal at the user. For this purpose, in this section, we propose a phase shift design at the LIS that maximizes the gain of the estimated composite downlink channel given by $\| \check{\mathbf{G}}^T \mathbf{\Phi} \check{\mathbf{H}}^T + \check{\mathbf{Z}}^T\|^2_{F}$. Due to the fixed precoding/beamforming strategy at the AP, we have selected the total channel gain as an optimization objective. To find the optimal phase shifts, $\mathbf{\Phi}^{\star}$, we formulate an optimization problem which can be expressed as
\begin{align}
    &\ \ \ \ \ \ \ \max_{\mathbf{\Phi}} \left\| \check{\mathbf{G}}^T \mathbf{\Phi} \check{\mathbf{H}}^T + \check{\mathbf{Z}}^T\right\|^2_{F} \label{opt1}\\
    & \textrm{s. t.} \ \ \ \ \ \mathbf{\Phi}_{l,l} = e^{j\phi_{l}}, \ \ l=1,\ldots,L.
\end{align}
Note that the optimization of phase shifts is performed at the AP, which is capable of managing the phase shifts at the LIS. The problem \eqref{opt1} is non-convex optimization problem since $\| \check{\mathbf{G}}^T \mathbf{\Phi} \check{\mathbf{H}}^T + \check{\mathbf{Z}}^T\|^2_{F}$ can be shown to be a non-concave function over $\mathbf{\Phi}$ \cite{zhang2019capacity}. Therefore, in order to solve the optimization problem, we rely on its approximation. For notational simplicity, let us assume that $\tilde{\mathbf{G}}=\check{\mathbf{G}}^T$, $\tilde{\mathbf{H}}=\check{\mathbf{H}}^T$ and $\tilde{\mathbf{Z}}=\check{\mathbf{Z}}^T$. Let $\mathbf{A}=\tilde{\mathbf{G}} \mathbf{\Phi} \tilde{\mathbf{H}} + \tilde{\mathbf{Z}}$, now we can write $\| \tilde{\mathbf{G}} \mathbf{\Phi} \tilde{\mathbf{H}} + \tilde{\mathbf{Z}}\|^2_{F}$ in \eqref{opt1} as
\begin{equation}\label{app1}
    \|\mathbf{A}\|^2_F = \sum_{i=1}^{N} \sum_{j=1}^{M} |a_{i,j}^{(l)}|^2, 
\end{equation}
where 
\begin{equation}
 \left|a_{i,j}^{(l)} \right|^2 = \left|\sum_{l=1}^{L}\tilde{g}_{i,l} e^{j\phi_l} \tilde{h}_{l,j} + \tilde{z}_{i,j}\right|^2. \label{aij}
 \end{equation}
 We can rewrite \eqref{aij} as 
 \begin{align}
  \left|a_{i,j}^{(l)} \right|^2&=\left(\sum_{l=1}^{L}\tilde{g}_{i,l} e^{j\phi_l} \tilde{h}_{l,j} + \tilde{z}_{i,j}\right)^{*} \left(\sum_{l=1}^{L}\tilde{g}_{i,l} e^{j\phi_l} \tilde{h}_{l,j} + \tilde{z}_{i,j}\right) \nonumber \\
 &=\left(\sum_{l=1}^{L}\tilde{g}_{i,l} e^{j\phi_l} \tilde{h}_{l,j}\right)^{*} \left(\sum_{l=1}^{L}\tilde{g}_{i,l} e^{j\phi_l} \tilde{h}_{l,j} \right) + |\tilde{z}_{i,j}|^2 + \nonumber \\
 & \ \ \ \ 2 \ \text{Re}\left\{ \tilde{z}_{i,j}^*\left(\sum_{l=1}^{L}\tilde{g}_{i,l} e^{j\phi_l} \tilde{h}_{l,j}\right) \right\}.\label{ps}
\end{align}
The first term in \eqref{ps} can be further expressed as 
\begin{align}
   &\hspace{-.3em}\left(\sum_{l=1}^{L}\tilde{g}_{i,l} e^{j\phi_l} \tilde{h}_{l,j}\right)^{*} \left(\sum_{l=1}^{L}\tilde{g}_{i,l} e^{j\phi_l} \tilde{h}_{l,j}\right)= & \label{ps2}\\
    &\sum_{l=1}^{L} \hspace{-.2em}\left|\tilde{g}_{i,l}\right|^2 \left|e^{j\phi_l}\right|^2 \left|\tilde{h}_{l,j}\right|^2 +  \sum_{l=1}^{L}\sum_{q\neq l}^{L} \left(\tilde{g}_{i,l}^* e^{-j\phi_l} \tilde{h}_{l,j}^* \tilde{g}_{i,q} e^{j\phi_q} \tilde{h}_{l,j}\right) & \nonumber
\end{align}
We note that the first term in \eqref{ps2} has no phase information because of the fact that $|e^{j\phi_{l}}|^2 =$ 1. On the other hand, the second term which represents the cross terms in \eqref{ps2} has negligible value as $2 \text{Re}\{ \tilde{z}_{i,j}^* (\sum_{l=1}^{L}\tilde{g}_{i,l} e^{j\phi_l} \tilde{h}_{l,j}) \} >> \sum_{l=1}^{L}\sum_{q\neq l}^{L} (\tilde{g}_{i,l}^* e^{-j\phi_l} \tilde{h}_{l,j}^* \tilde{g}_{i,q} e^{j\phi_q} \tilde{h}_{l,j})$. Thus, we ignore the first term in \eqref{ps} and other terms which are independent of phase shifts $\phi$ and we can rewrite our approximated total channel gain maximization optimization problem as 
\begin{align}
    &\ \ \ \ \ \ \ \max_{\phi_l} \ \ \sum_{i=1}^{N} \sum_{j=1}^{M} 2\text{Re}\left\{ \tilde{z}_{i,j}^*\left(\sum_{l=1}^{L}\tilde{g}_{i,l} e^{j\phi_l} \tilde{h}_{l,j}\right) \right\}. \label{opt2}
\end{align}
From \eqref{opt2}, we can obtain the phase of the $l^{\text{th}}$ element in the LIS, given by $\phi_l^{\star}=\mathbf{\Phi}_{l,l}^{\star}$, that maximizes the approximate version of the total channel gain as
\begin{equation}\label{fin}
    \phi_l^{\star}=\text{tan}^{-1} \left(\frac{\text{Im}\left\{ \sum_{i=1}^{N} \sum_{j=1}^{M}  \tilde{z}_{i,j} \tilde{g}_{i,l} \tilde{h}_{l,j} \right\}}{\text{Re}\left\{ \sum_{i=1}^{N} \sum_{j=1}^{M} \tilde{z}_{i,j} \tilde{g}_{i,l} \tilde{h}_{l,j} \right\}}\right).
\end{equation}
Since the fixed eigenmode transmit precoding and receive combining scheme is considered in this study, after obtaining $\mathbf{\Phi}^{\star}$ using \eqref{fin}, we compute the precoding/combining vectors using the composite channel matrix given by $\tilde{\mathbf{G}} \mathbf{\Phi}^{\star} \tilde{\mathbf{H}} + \tilde{\mathbf{Z}}$. It is also possible to perform a joint optimization of downlink precoding and LIS phase shift, however, the investigation of the joint optimization is out of the scope of this study.

\section{Numerical Results}\label{num_sec}

In this section, we evaluate the performance of our proposed two-stage channel estimation technique for the LIS assisted MIMO system. First, the performance of the proposed BAdVAMP based channel estimation algorithm is compared with its counterpart the BiGAMP algorithm  \cite{He2019CascadedCE}, in terms of normalized mean squared error (NMSE). Later, we present the achievable rate performance of the LIS assisted MIMO communication system using the optimized LIS phase shift design.

\textbf{MIMO channel parameters:} We use narrowband geometric channel model \eqref{eq1} to obtain the MIMO channels $\mathbf{H}$, $\mathbf{G}$ and $\mathbf{Z}$. The number of propagation paths in $\mathbf{Z}$ and $\mathbf{H}$ are set to $L_Z=L_H=64$. Due to highly correlated propagation conditions, the number of propagation paths in $\mathbf{G}$ is set to $L_G=8$ with $r_g=\text{rank}(\mathbf{G})=8$. Similar to \cite{BAdVAMP}, the singular values of $\mathbf{H}$ are selected to obtain a desired condition number $\kappa(\mathbf{H})$ while ensuring that $\mathbb{E}\left[ \| \mathbf{H}\|^2_F\right]=LM$. In \eqref{eq1}, angles of departure and arrival in both azimuth and elevation are distributed with the Laplacian distribution \cite{6717211}. For simplicity, the angular spreads (standard deviations) of these angles are assumed equal and set to $10^0$. Similar to $\mathbf{H}$, we also ensure that $\mathbb{E}\left[ \| \mathbf{G}\|^2_F\right]=NL$ and $\mathbb{E}\left[ \| \mathbf{Z}\|^2_F\right]=NM$. Throughout this section, the inter-element spacing at the user is assumed to be half-wavelength, whereas, we consider two cases of inter-element spacing at the AP with $d=\lambda/2$ and $d=4\lambda$.

\textbf{BAdVAMP initialization:} For initialization of the BAdVAMP algorithm, we draw $\mathbf{r}_{1,tr}^0$  and $\bm{\psi}^0_H$ from the i.i.d. $\mathcal{CN}(0,10)$ and $\mathcal{CN}(0,1)$ distributions, respectively. The initial value of $\gamma_{1,tr}^0$ is set to $10^{-3}$ and it is assumed that the statistical distribution of the matrix $\mathbf{D}$, i.e., $\bm{\theta}_d$, is known at the AP. Inner EM iterations are set to $\tau_{1,\text{max}}=1$ and $\tau_{2,\text{max}}=0$. The maximum number of iterations, $T_{\text{max}}$, for both BAdVAMP and BiGAMP schemes is fixed to 300, where the maximum number of restarts in BAdVAMP is set to 10.

\subsection{BAdVAMP versus BiGAMP channel estimation}

In this subsection, various experiments are carried out to compare the performance of the proposed BAdVAMP based channel estimation scheme with the BiGAMP channel estimation scheme \cite{He2019CascadedCE}. The NMSE performance comparison presented here is for the LIS channels $\mathbf{H}$ and $\mathbf{G}$ only. For the estimated matrix $\hat{\mathbf{H}}$ (before removing the ambiguities), the NMSE is given by 
\begin{equation}
    \text{NMSE}\left(\check{\mathbf{H}}\right)  = \frac{\|\mathbf{H} - \hat{\mathbf{H}}\mathbf{P}\|^2_F}{\|\mathbf{H}\|^2_F},
\end{equation}
where $\mathbf{P}$ denotes the generalized permutation matrix which perfectly removes the permutation, scalar and phase ambiguities. This perfect elimination of ambiguities is only considered in this subsection for comparison purposes.

\begin{figure}[!t]
	\centering
	\includegraphics[width=.94\linewidth]{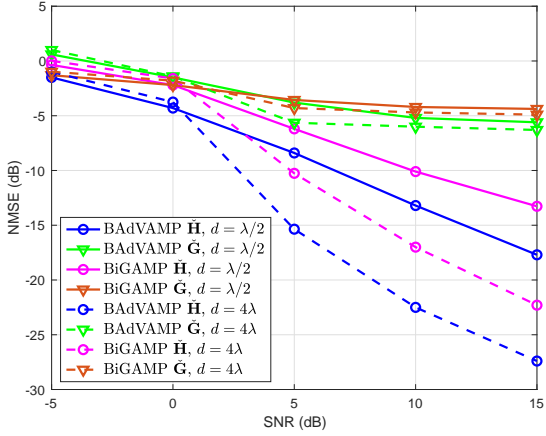}
	\caption{NMSE performance of estimated $\mathbf{H}$ and $\mathbf{G}$ against SNR values with $d=\{\lambda/2,4\lambda\}$ and $M=N=L=64$.}
	\label{fig11}
\end{figure}
In Fig. \ref{fig11}, we plot NMSE results of the LIS estimated channels $\mathbf{H}$ and $\mathbf{G}$ against various values of SNR for both BAdVAMP and BiGAMP based channel estimation schemes. Here, we set the total number of transmit antennas and receive antennas to $M=N=64$. The number of passive elements in LIS is $L=64$ and the length of the training sequences is $T_r=500$. The SNR is defined as $(1/\sigma^2_n)$, where $\sigma^2_n$ is the noise variance. The entries of the training matrix $\mathbf{X}_b$ follow the $\mathcal{CN}(0,1)$ distribution and then it is normalized, such that $\|\mathbf{X}_b\|^2_F=p_u$. The sparsity rate $K/N$ of the matrix $\mathbf{S}$ during the training stage is set to 0.1, meaning that each column of $\mathbf{S}$ consists of $K$ non-zero entries with $K=\ceil{0.1N}$, where $\ceil{\cdot}$ represents the ceiling function. The condition number of the LIS channel matrix $\mathbf{H}$ is set to $\kappa(\mathbf{H})=100$. In Fig. \ref{fig11}, two cases of inter-element spacing at the AP is considered: $d=\lambda/2$ and $d=4\lambda$. Fig. \ref{fig11} indicates that the proposed BAdVAMP based channel estimation technique estimates the LIS channels $\mathbf{H}$ and $\mathbf{G}$ with a higher quality as compared to the BiGAMP algorithm \cite{He2019CascadedCE}, especially at moderate to high SNR regimes. For the channel matrix $\mathbf{H}$, it can be seen that the NMSE performance gap between the two algorithms widens as the SNR increases. This higher quality estimation of the LIS channels is due to the fact that the VAMP part of the BAdVAMP algorithm is robust to ill-conditioned matrices as compared to the AMP based BiGAMP algorithm \cite{BAdVAMP}. Further, we note that the performance of both the schemes is similar for the estimated LIS channel $\mathbf{G}$.

\begin{figure}[!t]
	\centering
	\includegraphics[width=.94\linewidth]{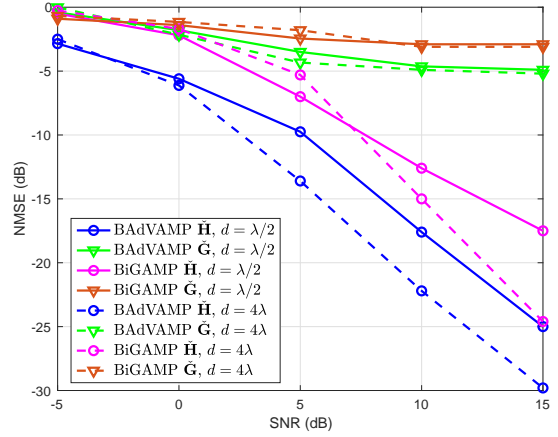}
	\caption{NMSE performance of estimated $\mathbf{H}$ and $\mathbf{G}$ against SNR values with $d=\{\lambda/2,4\lambda\}$ and $M=N=L=36$.}
	\label{fig22}
\end{figure}
A similar trend can be seen in Fig.~\ref{fig22} which demonstrates the NMSE performance against various SNR values with $M=N=L=36$. Here, the rest of the simulation parameters are the same as used in Fig. \ref{fig11}. The spatial correlation in $\mathbf{H}$ with $d=\lambda/2$ is higher than the case when $d=4\lambda$. Both Fig. \ref{fig11} and Fig. \ref{fig22} indicate that the NMSE performance improves when the channel matrix is spatially less correlated i.e., for $d=4\lambda$ case. However, in both the cases, BAdVAMP is preferable as it yields the estimated channels with higher quality than its counterpart BiGAMP. 

\begin{figure}[!t]
	\centering
	\includegraphics[width=.94\linewidth]{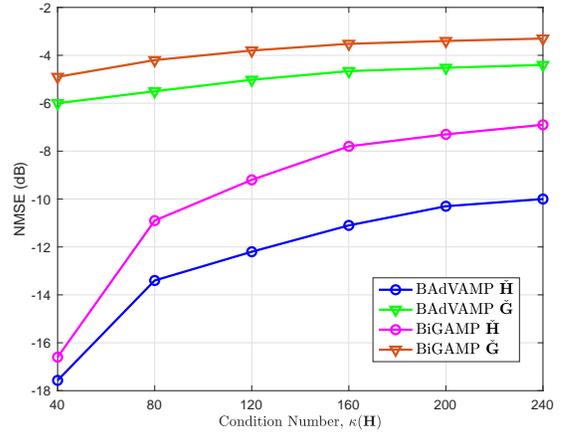}
	\caption{NMSE performance of estimated $\mathbf{H}$ and $\mathbf{G}$ against $\kappa(\mathbf{H})$ values with $M=N=L=64$.}
	\label{fig33}
\end{figure}
To examine the performance of the proposed scheme against different conditioning levels of the channel matrix $\mathbf{H}$, we plot NMSE performance against various $\kappa(\mathbf{H})$ values in Fig. \ref{fig33}. In this figure, we set $M=N=L=64$, SNR$=10$dB and $d=\lambda/2$. To induce conditioning in $\mathbf{H}$, we vary the condition number of the channel matrix $\mathbf{H}$, i.e., $\kappa(\mathbf{H})$, from $40$ to $240$. Note that as $\kappa(\mathbf{H})$ increases, the channel matrix $\mathbf{H}$ becomes more ill-conditioned. Fig. \ref{fig33} indicates the robustness of the proposed BAdVAMP scheme over the BiGAMP scheme when learning ill-conditioned channels. Since both the schemes suffer from higher ill-conditioning, we note that the performance degradation of the BAdVAMP scheme is with a slower rate than the BiGAMP scheme.  

\subsection{Achievable rate with BAdVAMP and optimized $\bm{\mathit{\Phi}}$}
Here, we evaluate the achievable rate performance of the LIS assisted MIMO system using the proposed LIS phase shift design presented in Section \ref{p_shift}. For the rest of the results presented in this section, we fix $N_s=2$, inter-element spacing at the AP to $d=\lambda/2$ and $p_u=1$. The value of the training length in the first stage of the estimation process is set to $64$, i.e., $T_d=64$ and $\sigma_e^2=\sigma_n^2M/(p_u+\sigma_n^2M)$. We do not assume that the ambiguities are eliminated perfectly but it is considered here that only the permutation ambiguity is removed from the estimated channels using the framework provided in Section \ref{recov}.
\begin{figure}[!t]
	\centering
	\includegraphics[width=.94\linewidth]{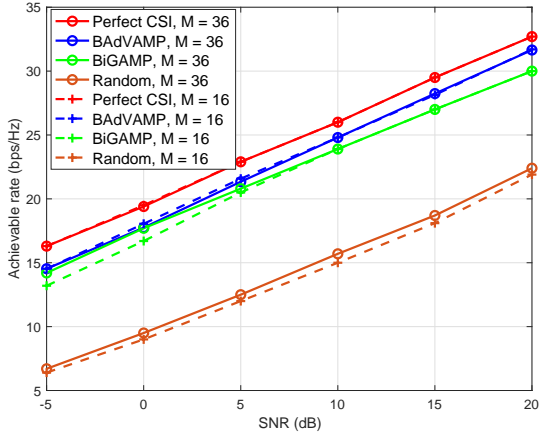}
	\caption{Achievable rate versus SNR performance with $M=N=\{36,16\}$, $L=36$ and $\kappa(\mathbf{H})=160$.}
	\label{fig44}
\end{figure}
Achievable rates of two different configurations are plotted in in Fig.~\ref{fig44}, where the cases considered are: i) $M=N=36$ and ii) $M=N=16$. Here, we use $L=36$, $T_r=800$ and $\kappa(\mathbf{H})=160$. Fig.~\ref{fig44} demonstrates the achievable rate, given by \eqref{cap}, for the proposed channel estimation and optimal phase shift design, labeled as `BAdVAMP'. We compare it with the scenario `Perfect CSI', where perfect CSI is assumed to be available at the AP and the associated LIS phase shift design is based on exhaustive search in problem \eqref{opt1}. For a comparison, we also include a scenario `Random', where random precoding/combining scheme and random LIS phase shifts are used. Finally, we also plot the achievable rate results with `BiGAMP' based channel estimation \cite{He2019CascadedCE} using our proposed LIS phase shift design as the phase shift optimization was not investigated in \cite{He2019CascadedCE}. 
\begin{figure}[!t]
	\centering
	\includegraphics[width=.94\linewidth]{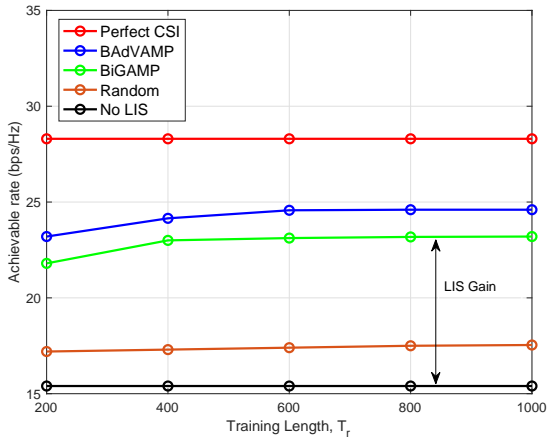}
	\caption{Achievable rate versus SNR performance with $M=N=L=64$, SNR$=10$ dB and $\kappa(\mathbf{H})=160$.}
	\label{fig55}
\end{figure}
Fig.~\ref{fig44} indicates that the proposed BAdVAMP scheme achieves better rate performance than the BiGAMP and random schemes. For example, at SNR$=20$ dB, the achievable rate gap between the proposed scheme and the BiGAMP scheme is nearly equal to 1.65 bps/Hz. More importantly, it can be seen that the rate performance with the proposed phase shift design is close to the perfect CSI case. Further, we note that the achievable rate performance at $M=36$ and $M=16$ are comparable. The reason for this trend is that we only use $N_s=2$ data streams, whereas, increasing the number of data streams will make the rate gap more apparent for these two antenna configurations.

Next, we evaluate the achievable rate performance against the training length, $T_r$ in Fig.~\ref{fig55}, where we set $M=N=L=64$, SNR $=10$ dB and $\kappa(\mathbf{H})=160$. Similar to the previous figures, we compare the achievable rate results with perfect CSI, BiGAMP and random schemes. However to gain more insight into the usefulness of LIS, here we also include a case where no LIS is present in the surrounding area. It can be seen in Fig.~\ref{fig55}, that initially the achievable rate increases with the increase in the training length for both BAdVAMP and BiGAMP schemes, however, at $T_r=600$, the rate performance converges and does not increases with $T_r$. This suggests that the training length of $600$ is the optimal choice for the given network settings. Fig.~\ref{fig55} also indicates that the proposed scheme performs better than the case when no LIS terminal is available to assist the MIMO communication. In fact, the rate performance with no LIS is even lower than the random (with LIS) case. This trend supports the idea of deploying the LIS terminal. 
\begin{figure}[!t]
	\centering
	\includegraphics[width=.94\linewidth]{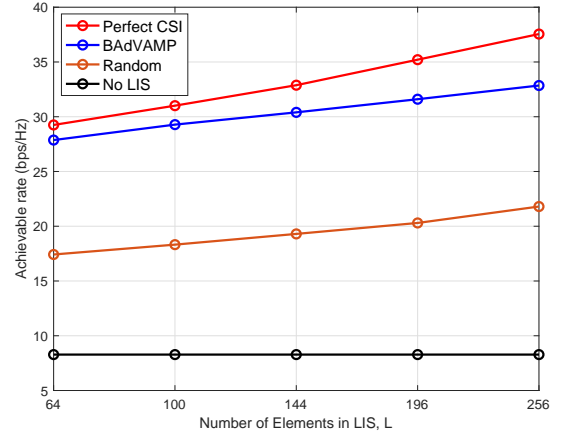}
	\caption{Achievable rate versus the number of LIS elements performance with $M=N=36$ and SNR$=10$ dB.}
	\label{fig66}
\end{figure}
Finally in Fig.~\ref{fig66}, we show the achievable rate performance by varying the number of passive elements, $L$, in the LIS from 64 to 256. Here, we set $M=N=36$, SNR$=10$ dB and $\kappa(\mathbf{H})=100$. It can be observed from Fig.~\ref{fig66} that the rate improves by increasing the number of passive elements in the LIS. 

\section{Conclusions}\label{conc}
In this paper, we have proposed a two stage channel estimation method for the LIS assisted MIMO communication system, where the direct channel between the AP and user is estimated in the first stage and the BAdVAMP algorithm is used to estimate LIS channels in the second stage. We also presented a phase shift design for the LIS which approximately maximizes the channel gain at the user. Through numerical simulations, we show the effectiveness of the proposed channel estimation method and phase shift design.

There are some key observations in this study which needs to be highlighted. First of all, we can see that the BAdVAMP algorithm outperforms the BiGAMP algorithm in learning both ill-conditioned and well-conditioned channel matrices. We also observed that LIS is beneficial for the MIMO configurations as it improves the achievable rate performance compared to the conventional MIMO system having no LIS terminal. For future work, it will be interesting to investigate the joint MIMO precoding/combining and LIS phase shift optimization. It will also be useful to optimize the training length such that a balance is achieved between the channel estimation quality and training overhead. Furthermore, channel estimation strategies should also be investigated in scenarios where multiple reflecting surfaces are present.

\bibliographystyle{IEEEtran}
\bibliography{ref}

\end{document}